\documentstyle[amscd,amssymb,verbatim,12pt]{amsart}
\def\dspace{\baselineskip=0.3 in}
\begin{document}
\dspace
\title[Riccion as a cosmic ......]{Riccion as a cosmic dark matter candidate
  and late cosmic acceleration}

\author[S.K.Srivastava and K.P.Sinha]%
        {    }

\maketitle

\centerline{\bf S.K.Srivastava }

\centerline{ Department of Mathematics, North Eastern Hill University,}

\centerline{ NEHU Campus,Shillong - 793022 ( INDIA ) }

\centerline{e-mail:srivastava@@nehu.ac.in  }

\centerline{and }

\centerline{\bf K.P. Sinha}

\centerline{ INSA Honorary Scientist, Department of Physics,}

\centerline{ Indian Institute of Science , Bangalore - 560012 ( INDIA ) .}

\centerline{e-mail: kpsinha@@gmail.com }

\vspace{1cm}

\centerline{\bf Abstract}

In the past few years, a posibility is investigated, where curvature
itself behaves as a source of dark energy. So, it is natural to think whether
curvature can produce dark matter too. It is found that,
at classical level, higher-derivative gravity yields curvature inspired
particles namely {\em riccions}\cite{rr}. Here, it is probed whether {\em
  riccion} can be 
a possible source of dark matter. Further, it is found that the late universe
accelerates. Here, it is interesting to see that acceleration is obtained from
curvature without using any dark energy source of exotic matter.



\vspace{1.5cm}

By the end of the last century, cosmology got revolutionized due to Supernova
observations at low red-shift, pointing towards very late cosmic acceleration
and $73\%$ content of the present universe as dark energy (DE)\cite{sjp, agr1,
  jlt, pg, bps, sjp2}. Above
all these experiments, WMAP data provide estimates of parameters of standard
models of cosmology with unprecedented accuracy \cite{clb,abl}. These data
suggest  $73\%$ content of the universe in the form of DE,
$23\%$ in the form of non-baryonic dark matter (DM) and the rest  $4\%$  in the form
of baryonic matter as well as radiation. But identity of dark energy and dark
matter is still in dark.

For DE, many theoretical models were proposed in the past few years
taking exotic matter ( quintessence, tachyon , phantom and k-essence )as a
source \cite[for detailed review]{scj}. Later on, higher-order curvature terms
or functions of curvature were taken as gravitational alternative of dark
energy \cite{snsd}.  In contrast to this approach of taking non-linear
curvature terms as DE lagrangian, in \cite{sksde}, a different approach is adapted where
non-linear curvature terms are {\em not} taken as DE lagrangian. Rather,in \cite{sksde}, DE terms
emerge spontaneously in the resulting Friedmann equations (giving cosmic
dynamics), if non-linear curvature terms are added to Einstein - Hilbert term
in the gravitational action.

Here, we address to the identity of DM. In principle, DM can be hot as well as
cold. The possibility of hot dark matter is ruled out by WMAP data due to
re-organization of the universe at red-shift $ z \simeq 20$ . So, in the
present universe, dominant component of DM is {\em cold} and {\em
  non-baryonic} \cite{abl}. It is further
characterized as a {\em pressureless} cosmic fluid. In the search of cold dark
matter (CDM) candidate, some attempts have been made from time to time in the
past. In \cite{jc}, supersymmetric dark matter is suggested with particle mass
$\lesssim 500$ GeV. In \cite{tn}, wino-like neutrilo, with mass of the order $
100$ GeV, is suggested as a CDM candidate. In \cite{ft}, it is shown that
pseudo-Nambu-Godostone boson can be a possible source of DE and its
supersymmetric partner can give dark matter. The possibility of Kaluza-Klein
dark matter is discussed in \cite{gs, gs2, hcc, dh, gb, fb}.

In this letter, there is a digression from so far attempts to identify CDM. It
is mentioned above that DE can be obtained from the gravitational sector also
using higher-order curvature terms without resorting to exotic matter. So, it
is natural to think whether it is possible to obtain CDM too from gravity with
non-linear curvature terms.

In \cite{sksr, sksp, sksj, sksi, sksm, sksil, sksij, sksmp, sksijm, skspra, rr}, it is shown that when higher-order curvature terms are added
to Einstein-Hilbert lagrangian, Ricci scalar manifests itself in {\em dual} manner
(i) as a geometrical field and (ii) as a physical field. Its physical aspect
is given by a particle called {\em riccion} with (mass)$^2$ inversely
proportional to the gravitational constant $G$. Higher-drivative gravity has
unitarity problem at quantum level, but, at classical level, it does not face
any 
problem if coupling constants are taken properly. In what follows as well as in
\cite{sksr,  sksp, sksj, sksi, sksm, sksil, sksij, sksmp, sksijm, skspra, rr}, it is found that the scalar field, namely {\em riccion}, is
{\em different} from the {\em scalar mode of graviton} as {\em graviton} is
massless, whereas {\em riccion} is a particle with mass proportional to
inverse of square root of the gravitational constant.   A detailed discussion on the difference between {\em riccion} and
{\em   graviton } is given in \cite[Appendix A]{rr}. In what follows, to get
curvature inspired DM, we probe an answer to the question,``Is riccion a
 cosmic dark matter candidate too ?'' 

Here, natural units ($k_B = {\hbar} = c = 1$) are used with GeV as a
fundamental  unit with $k_B$ being the Boltzman constant, ${\hbar}$ being the
Planck's constant $h$ divide by $2\pi$ and $c$ being the speed of light. 

We take $R^2$-gravity being the simplest higher-derivative theory of gravity
with the 
action 
$$ S_g = \int {d^4x} \sqrt{-g} [\frac{R}{16\pi G} + \alpha R^2], \eqno(1)$$
where $\alpha$ is a dimensionless coupling constant. 
This action yields the gravitational field equations
$$\frac{1}{16\pi G}[R_{\mu\nu} - \frac{1}{2} g_{\mu\nu} R] + 2 \alpha
[R_{;\mu\nu} - g_{\mu\nu} {\Box} R + R R_{\mu\nu}] -  \frac{1}{2} g_{\mu\nu}
R^2 = 0,   \eqno(2)$$
where semicolon (;) stands for the covariant derivative and the operator ${\Box} = ({1}/{\sqrt{-g}})
\partial_{\mu}[\sqrt{-g} g^{\mu\nu} \partial_{\nu}].$
Trace of these field equations are obtained as
$$ {\Box} R + \frac{1}{96\pi G\alpha} R = 0  \eqno(3)$$
with $\alpha > 0$ to avoid the ghost problem.

This is the Klein-Gordon equation for $R$. It
 shows that when higher- order terms of curvature are added to
Einstein- Hilbert lagrangian $R/{16\pi G}$ in $S_g, R$ behaves as a
physical field also \cite{sksr,  sksp, sksj, sksi, sksm, sksil, sksij, sksmp,
 sksijm, skspra, rr}.

$R$ has mass dimension 2, being combination of of second order derivative and
squares of the first order derivatives of $g_{\mu\nu}$ (which is
dimensionless).

In a scalar field theory, scalars satisfying the Klein-Gordon equation have
mass dimension 1. So, to have consistency with other scalar fields, we
multiply (3) by $\eta$, which is a constant having (mass)$^{-1}$ dimension and
measured in GeV$^{-1}$. $\eta R$ is recognized as ${\tilde R}$, which is
called {\em riccion} having mass dimension 1. It satifies the equation
$$ {\Box} {\tilde R} + m^2 {\tilde R} = 0  \eqno(4)$$
with
$$ m^2 = \frac{1}{96\pi G\alpha}. \eqno(5)$$

If $\tilde R$ is a physical field (as mentioned above), there should be an
action $S_{\tilde R}$ yielding (4) on using its variation with respect to
$\tilde R$. Here, we find $S_{\tilde R}$ as given below. According to these
requirements for $S_{\tilde R}$, we have  \cite{sksr,  sksp,  sksj, sksi, sksm,
  sksil, sksij, sksmp, sksijm, skspra, rr}
$$\delta S_{\tilde R} = - \int {d^4 x}\sqrt{-g} [{\Box} {\tilde R} + m^2 {\tilde R}]
\delta {\tilde R}, $$
subject to the condition $\delta S_{\tilde R}/\delta {\tilde R} = 0.$ This
equation reduces to
$$\delta S_{\tilde R} =  \int {d^4 x}\sqrt{-g} [g^{\mu\nu} \partial_{\mu} {\tilde R}
\partial_{\nu} (\delta {\tilde R})- m^2 {\tilde R}\delta {\tilde R}]
, $$
which is {\em covariant}. So, using the {\em principle of equivalence}, we have
$$\delta S_{\tilde R} =  \int {d^4 X} [\eta^{\mu\nu} \partial_{\mu} {\tilde R}(X)
\partial_{\nu} (\delta {\tilde R})(X) - m^2 {\tilde R}(X)\delta {\tilde R}(X)]
\eqno(6a)$$
in a locally inertial co-ordinate system (X) having
$$ g_{\mu\nu} (X) = \eta_{\mu\nu} \quad{\rm and}\quad \Gamma^{\mu}_{\nu\rho}(X)
= 0. $$

Now, (6a) integrates to
\begin{eqnarray*}
 S_{\tilde R} &=& \frac{1}{2} \int {d^4 X} [\eta^{\mu\nu} \partial_{\mu} {\tilde R}(X)
\partial_{\nu}{\tilde R}(X) - m^2 {\tilde R}^2(X)]\\&=& \frac{1}{2} \int {d^4
  x}\sqrt{-g} [g^{\mu\nu} \partial_{\mu} {\tilde R} \partial_{\nu}{\tilde R} - m^2 {\tilde R}^2]
\end{eqnarray*}
\vspace{ - 1.8cm}
\begin{flushright}
(6b)
\end{flushright}
due to {\em principle of equivalence} and {\em principle of covariance}
\cite{rr}.

Like other physical fields, energy-momentum tensor components for ${\tilde R}$
can be obtained varying $S_{\tilde R}$ with respect to $g^{\mu\nu}$. This
variation is obtained from (6b) as
\begin{eqnarray*}
 \delta S_{\tilde R} &=&  \int {d^4 }\sqrt{-g} \Big[ \frac{1}{2} \delta g^{\mu\nu}
 \partial_{\mu} {\tilde R}\partial_{\nu}{\tilde R} + g^{\mu\nu}
 \partial_{\mu} {\tilde R}\partial_{\nu}\delta{\tilde R}- \frac{1}{2} m^2 {\tilde
 R}\delta{\tilde R} \\&&- \frac{1}{2} g_{\mu\nu} \delta
 g^{\mu\nu}\Big\{\frac{1}{2} \partial^{\rho} {\tilde R} \partial_{\rho}{\tilde
 R} - m^2 {\tilde R}^2 \Big\} \Big] \\&=& \int {d^4 } \sqrt{-g}\Big[ \frac{1}{2} \delta g^{\mu\nu}
 \partial_{\mu} {\tilde R}\partial_{\nu}{\tilde R} - {\Box} {\tilde R}\delta{\tilde R}- m^2 {\tilde
 R}\delta{\tilde R} \\&&- \frac{1}{2} g_{\mu\nu} \delta
 g^{\mu\nu}\Big\{\frac{1}{2} \partial^{\rho} {\tilde R} \partial_{\rho}{\tilde
 R} - \frac{1}{2} m^2 {\tilde R}^2 \Big\} \Big].
\end{eqnarray*}
\vspace{ - 1.8cm}
\begin{flushright}
(7a)
\end{flushright}

On using (4) in (7a),
$$\delta S_{\tilde R} =  \int {d^4 x}\sqrt{-g} \Big[ \frac{1}{2}
 \partial_{\mu} {\tilde R}\partial_{\nu}{\tilde R} - \frac{1}{2} g_{\mu\nu}\Big\{\frac{1}{2} \partial^{\rho} {\tilde R} \partial_{\rho}{\tilde
 R} - \frac{1}{2} m^2 {\tilde R}^2 \Big\} \Big]\delta g^{\mu\nu}, \eqno(7b)$$
which yields
\begin{eqnarray*}
 T_{\mu\nu}^{({\tilde R})} &=& \frac{2}{\sqrt{- g}} \frac{\delta S_{\tilde R}}{
 \delta g^{\mu\nu}} \\&=& \partial_{\mu} {\tilde R}\partial_{\nu}{\tilde R} - g_{\mu\nu}\Big\{\frac{1}{2} \partial^{\rho} {\tilde R} \partial_{\rho}{\tilde
 R} - \frac{1}{2} m^2 {\tilde R}^2 \Big\}.
\end{eqnarray*}
\vspace{ - 1.8cm}
\begin{flushright}
(8)
\end{flushright}

Experimental observations \cite{adm, pd20, ael, am20, sh20} support spatially homogeneous and
isotropic cosmological
model for the universe, given by the metric
$$ ds^2 = dt^2 - a^2(t) [dx^2 + dy^2 + dz^2] , \eqno(9)$$
where $a(t)$ is the scale factor. So, due to homogeneity of space-time (9), ${\tilde R}$ ,being $\eta R$,depends on
cosmic time t only. 

In this space-time, {\em  energy density} and {\em
  pressure } of the cosmic fluid, constituted by spinless riccions
(represented by ${\tilde R}$ scalars)  are obtained from (8) as
$$ \rho^{({\tilde R})} = T^0_0 = \frac{1}{2} {\dot {\tilde R}}^2 + \frac{1}{2} m^2 {\tilde
  R}^2 \eqno(10)$$
and
$$ p^{({\tilde R})} = - T^1_1 = \frac{1}{2} {\dot {\tilde R}}^2 - \frac{1}{2} m^2 {\tilde
  R}^2 .\eqno(11)$$

(10) and (11) satisfy the conservation equation
$$ {\dot \rho^{({\tilde R})}} +  3 \frac{\dot a}{a} (\rho^{({\tilde R})}  +
p^{({\tilde R})}
) = 0 . \eqno(12)$$

Correctness of (12) can be verified easily. On using
(10) and (11) in (12), it is obtained that
$$ {\ddot {\tilde R}} + 3 \frac{\dot a}{a}{\dot {\tilde R}} + m^2 {\tilde R}^2
= 0,\eqno(13)$$
which is (4) (derived above from the action (1)), giving field equation for ${\tilde R}$ or riccion equation of
motion in the space-time (9) using (10) and 
$$ {\dot {\tilde R}}^2 = \rho^{({\tilde R})} + p^{({\tilde R})}$$
obtained from (10) and (11).

From (5), we find riccion as a massive particle with very high mass. So, it is
reasonable to think for riccions to behave like a gas of heavy massive and
very weakily interacting particles with negligibly small velocity distribution. So, these
particles are non-relativistic. Moreover, in the late universe, temperature is
low. Pressure density , due to non-relativistic particles , are obtained as \cite{rdt}
$$p^{({\tilde R})} = (m/2\pi)^{3/2} T^{5/2} exp[- m/T] \simeq  0  \eqno(14)$$
using Bose-Einstein distribution. It yields $w = p^{({\tilde R})}/ \rho^{({\tilde R})} = 0$. Here, $m$ is given
by (5). Moreover, $T \leq T_* = T_0  \Big({a_0}/{a_*} \Big) = 3.44 \times
10^{-13}{\rm   GeV} $ as it is found below that riccion fluid, behaving as CDM, dominates
at the red-shift $z_* = 0.46$ and ${a_0}/{a_*} = 1.46.$ from (30) given
below. Here $ T_0 = 2.73^0K = 2.35 \times 10^{-13}{\rm   GeV} $ is the present
temperature of the background radiation in the universe.

Though, even visible matter is pressureless with equation of state parameter
$w = 0$, riccion fluid can not behave as visible matter. It is because visible
matter is baryonic, whereas the riccion fluid, originating from  curvature, is
non-baryonic. So, it is possible for riccion to be a source of CDM.

Further, (11) and (14) imply
$$  \frac{1}{2} {\dot {\tilde R}}^2 = \frac{1}{2} m^2 {\tilde
  R}^2 .\eqno(15)$$

Connecting (10) and (15), it is obtaind that
$$ \rho^{({\tilde R})} = m^2 {\tilde   R}^2 \eqno(16)$$

Moreover, using (14) and (16) in (12) and integrating, it is obtained that
$$\rho^{({\tilde R})} = m^2 {\tilde   R}^2 = \frac{A}{a^3} , \eqno(17)$$

According to WMAP1 \cite{abl}, current value of CDM density $\rho_{\rm dm} $ is
$\rho^0_{\rm dm} = 0.23 \rho^0_{\rm cr}$, where
$\rho^0_{\rm cr} = 3 H_0^2/{8\pi  G}, H_0 = 100{\rm h} {\rm km/Mpc sec} = 2.33
\times 0.68 \times 10^{-42} {\rm GeV}$ using ${\rm h} = 0.68$ ( a value having
the maximum likelihood ). So, $\rho^0_{\rm dm }= 0.69 \times 10^{-47}
{\rm GeV}^4$ .

Using WMAP1 results, integration constant $A$ is evaluated as $A = 0.23
\rho^0_{\rm cr} a_0^3$, so from (17)
$$ R = \frac{\tilde R}{\eta} = \pm \sqrt{0.23 \rho^0_{\rm cr}}
\frac{a_0^{3/2}}{\eta m a^{3/2}} . \eqno(18)$$

In the space-time, given by the metric (9)
$$ R = 6 \Big[\frac{\ddot a}{a} + \Big(\frac{\dot a}{a} \Big)^2
\Big]. \eqno(19)$$

From (18) and (19), we obtain
$$ \frac{\ddot a}{a} + \Big(\frac{\dot a}{a} \Big)^2
 = \pm \sqrt{0.23 \rho^0_{\rm cr}}
\frac{a_0^{3/2}}{6 \eta m a^{3/2}}  . \eqno(20)$$

The first integral of (20) is obtained as
$$ \Big(\frac{\dot a}{a} \Big)^2 = \frac{C^2}{a^4} \pm \frac{2 \sqrt{0.23
    \rho^0_{\rm cr}}}{15 m \eta } \Big(\frac{a_0}{ a} \Big)^{3/2}  \eqno(21)$$
where $C$ is an integration constant. This is the Friedmann equation giving
 dynamics of the universe.

If we take $(-)$ sign in (18), ${\dot a}/{a}$ is complex, when
$$ \frac{C^2}{a^4} < \frac{2 \sqrt{0.23
    \rho^0_{\rm cr}}}{15 m \eta } \Big(\frac{a_0}{ a} \Big)^{3/2}. $$ 
So, from (21), we have
$$ \Big(\frac{\dot a}{a} \Big)^2 = \frac{C^2}{a^4} + \frac{2 \sqrt{0.23
    \rho^0_{\rm cr}}}{15 m \eta } \Big(\frac{a_0}{ a} \Big)^{3/2} . \eqno(22)$$

In case, 
$$ \frac{C^2}{a^4} > \frac{2 \sqrt{0.23
    \rho^0_{\rm cr}}}{15 m \eta } \Big(\frac{a_0}{ a} \Big)^{3/2}, \eqno(23)$$
(22) reduces to
$$ \Big(\frac{\dot a}{a} \Big)^2 \simeq \frac{C^2}{a^4} . \eqno(24)$$
This equation is integrated to
$$ a(t) = [D + 2 C t]^{1/2} \eqno(25)$$
showing deceleration as it yields ${\ddot a} < 0.$ Here $D$ is an integration
constant. 

When
$$ \frac{C^2}{a^4} < \frac{2 \sqrt{0.23
    \rho^0_{\rm cr}}}{15 m \eta } \Big(\frac{a_0}{ a} \Big)^{3/2}, \eqno(26)$$
(22) is approximated to
$$ \Big(\frac{\dot a}{a} \Big)^2 \simeq  \frac{2 \sqrt{0.23
    \rho^0_{\rm cr}}}{15 m \eta } \Big(\frac{a_0}{ a} \Big)^{3/2} . \eqno(27)$$
(27) yields the solution
$$ a(t) = \Big[E + 0.75 \sqrt{\frac{2 \sqrt{0.23
    \rho^0_{\rm cr}}}{15 m \eta }} {a_0}^{3/4} t\Big]^{4/3} \eqno(28)$$
showing acceleration as it yields ${\ddot a} > 0.$ Here $E$ is an integration
constant. 

Thus, it is obtained that universe decelerates when the inequality (23) holds
and accelerates  when the inequality (26) holds. It means that transition from 
deceleration to acceleration takes place at $ a = a_*,$ where $a_* = a(t_*)$
with $t_*$ being the transition time. So, at $t = t_*$,
$$ \frac{C^2}{a^4_*} = \frac{2 \sqrt{0.23
    \rho^0_{\rm cr}}}{15 m \eta } \Big(\frac{a_0}{ a_*} \Big)^{3/2}, \eqno(29)$$

16 Type Supernova observations \cite{agr} have conclusive evidence that, in
the late universe, acceleration begins at small red-shift $z_* \simeq 0.46$
with a jerk giving a transition from deceleration to acceleration.

So, we have
$$ \frac{a_0}{a_*} = 1 + z_* = 1.46 , \eqno(30)$$
where $a_0 = a(t_0)$ with $t_0 = 13.6 {\rm Gyr} = 6.6 \times 10^{41} {\rm
  GeV}^{-1}$ being the present time.

Connecting (29) and (30), $C$ is pbtained as
$$C = 1.33 a_*^2 \sqrt{\frac{2 \sqrt{0.23
    \rho^0_{\rm cr}}}{15 m\eta }} .$$
Using this value of $C$ and $a = a_*$ at $t = t_*$, (25) looks like
$$ a(t) = a_* \Big[ 1 + 1.33 a_*^2 \sqrt{\frac{2 \sqrt{0.23
    \rho^0_{\rm cr}}}{15 m \eta }} (t_* - t)\Big]^{1/2}  \eqno(31)$$
showing deceleration when $t < t_*.$

Moreover, using (30) in (28), it is obtained that
$$ a(t) = a_* \Big[ 1 + 0.996 \sqrt{\frac{2 \sqrt{0.23
    \rho^0_{\rm cr}}}{15 m \eta }} (t - t_*)\Big]^{3/4}  \eqno(32)$$
showing acceleration when $t > t_*.$ Using $a_0 = a(t_0)$ in (32), $t_*$ is
    obtained as
$$ t_* = t_0 - 0.65 \Big[ \sqrt{\frac{2 \sqrt{0.23
    \rho^0_{\rm cr}}}{15 m \eta }} \Big]^{-1}. \eqno(33)$$

In what follows, results are summarized. According to above investigations, 
$R^2$-gravity suggests that Ricci scalar manifests itself as a physical field
too in addition to its usual nature as a geometrical object. The dual nature
of $R$ has been found in the past in \cite{sksr, sksp, sksj, sksi, sksm, sksil, sksij, sksmp, sksijm, skspra, rr} and its various
consequences are discussed there. Manifestation of the physical aspect of the
Ricci scalar is realized through the scalar ${\tilde R}$ representing the
riccion. Thus the Ricci
scalar behaves simultaneously as a geometrical as well as a physical field
given by riccion. It is found that riccion is a heavy massive particle. So, it
is natural to think these particles to constitue a gas of heavy particles
interating very weakly with each other and moving with negligible speed. Such
a gas is  pressureless. Moreover, originating from curvature, riccions are
non-baryonic. So, it is obtained that riccions constitute pressueless
non-baryonic CDM. Further, it is
interesting to see that riccion CDM causes acceleration in the late universe.

Thus, we get acceleration in the late universe from $R^2$-gravity 
without taking any exotic matter as a dark energy source.
In \cite{om}, late cosmic acceleration is obtained modifying gravity with
inverse powers of linear combination of curvature invariants without taking
account of physical role of Ricci scalar manifested by riccion.

\end{document}